# An Ontological AI-and-Law Framework for the Autonomous Levels of AI Legal Reasoning

**Dr. Lance B. Eliot**
Chief AI Scientist, Techbruim; Fellow, CodeX: Stanford Center for Legal Informatics
Stanford, California, USA

**Abstract**

A framework is proposed that seeks to identify and establish a set of robust autonomous levels articulating the realm of Artificial Intelligence and Legal Reasoning (AILR). Doing so provides a sound and parsimonious basis for being able to assess progress in the application of AI to the law, and can be utilized by scholars in academic pursuits of AI legal reasoning, along with being used by law practitioners and legal professionals in gauging how advances in AI are aiding the practice of law and the realization of aspirational versus achieved results. A set of seven levels of autonomy for AI and Legal Reasoning are meticulously proffered and mindfully discussed.

## 1 Background and Context

Interest in applying Artificial Intelligence (AI) to the law has been existent since the early days of AI research [7] [15] [26] [31], having been variously attempted even during the initial formulation of computer-based AI capabilities, and continues earnestly today with ongoing efforts in a multitude of forums, including by AI development labs, by law scholars, by progressive law practices, by legal systems providers, and the like [8] [32] [45].

The field of AI and law has been populated over the years with numerous experimental systems that set new ground and inched forward progress in applying AI to the legal domain, notably TAXMAN [47], HYPO [2], CATO [6], and others [3] [54] have provided a foundation for incremental advancements. Despite these notable accomplishments, there is still much unexplored and unattained in terms of applying AI to the law. As eloquently phrased in the 1950s by Supreme Court Justice William Douglas [61]: "The law is not a series of calculating machines where answers come tumbling out when the right levers are pushed." His insightful remarks were true then and remain still true to this day.

At the core of the law is the human cognitive aspects for legal reasoning [62], and it is presumed that legal reasoning ultimately underlies all the salient acts of studying, conveying, and undertaking the practice of law [35] [36]. As such, AI as applied to the law is predominantly about the nature of legal reasoning and how this cognitive act can be undertaken by a computer-based system. Aptly stated by Ghosh [33]: "AI & Law is a subfield of AI research that focuses on designing computer programs, or computational models, that perform or simulate legal reasoning. In other words, AI & Law is the field of modeling computationally the legal reasoning for the purpose of building tools for legal practice."

### 1.1 Defining Automation versus Autonomy

It is customary in the legal reasoning context to divide the use of AI and computer-based systems into two focuses, one being the application of *automation* to the act of legal reasoning, which primarily then serves as an adjunct or augmentation to human legal reasoning efforts, and the other being the goal of achieving *autonomous* legal reasoning that consists of computer-based systems able to perform legal reasoning unaided by human legal reasoners and that can operate autonomously with respect to the practice of law. As per Galdon [30]: "Automation is defined as a system with a limited set of pre-programmed supervised tasks on behalf of the user. Autonomy, on the other hand, is defined as a technology designed to carry out a user's goals without supervision with the capability of learning and changing over time."

Law practices and legal professionals routinely today make use of *automation* in the performance of their needed tasks involving legal activities. A modern-day law office might use e-Discovery software as part of their case discovery pursuits and be perhaps crafting new contracts via the use of an online cloud-based service that pieces together prior contracts from a corpus established to enable reuse. Generally, the appropriate adoption of law-related computer-based systems has significantly aided lawyers and legal staff in un-dertaking their efforts, oftentimes cited as boosting efficiency and effectiveness accordingly. The automation being used for these purposes is not considered autonomous as yet, though advancements in these systems are being fostered by infusing AI capabilities to someday achieve autonomous operation.

A significant body of research exists on attempts to clarify what autonomy or autonomous operations consist of [42] [43]. There is much debate regarding the particulars of autonomy and different viewpoints ascribe differing qualities to the matter.

For example, Sifakis [60] defines that "autonomy is the capacity of an agent to achieve a set of coordinated goals by its own means (without human intervention) adapting to environment variations. It combines five complementary aspects: Perception e.g. interpretation of stimuli, removing ambiguity/vagueness from complex input data and determining relevant information; Reflection e.g. building/updating a faithful environment run-time model; Goal management e.g. choosing among possible goals the most appropriate ones for a given configuration of the environment model; Planning to achieve chosen goals; Self-adaptation e.g. the ability to adjust behavior through learning and reasoning and to change dynamically the goal management and planning processes."

Rather than weighing in herein on trying to pinpoint what autonomy entails per se, it is sufficient for the purposes of this discussion to consider that an *autonomous* computer-based system in this context would be one that can perform legal reasoning on its own, doing so without the aid of a human, and essentially perform legal reasoning that is on par with that of a human versed in legal reasoning [25] [26] [28].

Furthermore, it is prudent to refrain from discussing herein any indication about the AI techniques and technologies that might be required or employed to autonomously undertake the legal reasoning task, since doing so would tend to muddle the matter of concern. In essence, AI techniques and technologies are in a continual state of flux, being adjusted, refined, and at times formed a new, and the discussion herein might inadvertently get mired in AI that is known today but that might very well be improved or advanced tomorrow.

An exemplar would be the case of today's versions of Machine Learning (ML), and that Sifakis [60] emphasizes: "A main conclusion is that autonomy should be associated with functionality and not with specific techniques. Machine learning is essential for removing ambiguity from complex stimuli and coping with uncertainty of unpredictable environments. Nonetheless, it can be used to meet only a small portion of the needs implied by autonomous system design."

## 1.2 Establishing the Framework

The purpose of this paper is to identify and establish a set of automation and autonomous levels that can be applied to the law and therefore would articulate a framework for aiding and bolstering the realm of Artificial Intelligence (AI) and Legal Reasoning (AILR). There does not exist today an established taxonomy or framework that provides a set of autonomous levels for AILR [13] [26] [51].

This crucial point comes as a surprise to some that assumed or presumed that such a framework already existed. The lack of an established framework could be argued as an omission that has at times allowed for specious claims about what AI systems can do in the case of AILR and permitted misleading and at times outright false assertions by vendors or others, of which will likely inexorably further worsen as AI is increasingly infused into computer-based legal systems into the future.

Consider then these bases for justifying the formulation and promulgation of a bona fide set of autonomous levels for instituting a viable apples-to-apples depiction of the act of practicing law and the embodiment of AI-powered legal reasoning:

- Vendor offerings could be rated as to what level of automation or autonomy their wares truly provide, overcoming vacuous and otherwise unsubstantiated claims, allowing for easier and fair game comparisons.

- Law practices seeking to acquire or make use of legal systems would readily have the means to gauge what capabilities the automation or autonomy provides in those systems, and knowingly ascertain what they are getting and how to best implement such systems in their practices.

- Lawyers would be better informed as to the capabilities of AI-enabled legal systems, along with being able to assess the progress of automation that might serve either as an augmentation to their efforts or could potentially be an autonomous replacement for their efforts.

- Researchers and scholars would be able to ascertain what progress is being made in applying AI to the law, showcasing aspects that require further research and advancement and rely upon a validated framework as a barometer measuring the totality of the state of AILR.

- And so on.

In recap, there seems ample rationalization for putting in place an acknowledged and practical set of autonomous levels for AILR that would, therefore, provide a sound and robust basis for being able to assess progress in the application of AI to the law.

Accordingly, a framework consisting of a set of seven levels of autonomy for AI Legal Reasoning is proffered in this paper, accompanied by a carefully elucidated explanation that explicitly states the foundations used to formulate the framework. By including essential underpinnings, it is hoped that readers will be thusly cognizant of why the structure is shaped as so proposed (else it might seem haphazard or enigmatic as to the basis employed), and will also ensure a kind of open access for those that desire to refine or otherwise augment the draft framework.

## 2. Levels of Autonomy (LoA) Approaches

The notion of crafting a set of Levels of Autonomy (LoA) is not unheard of, indeed, there have been many earnest attempts and equally varied outcomes that have occurred over time to derive LoA's, as will be briefly examined next.

One perspective is that a generic LoA should be formulated and then applied "as is" to any domain seeking to embrace a set of levels of autonomous operations within that specific realm. Others argue that each domain dictates a tailored assessment of what an appropriate LoA ought to look like to adequately meet the needs of that sphere or discipline, and thus by implication that a wholly brand-new set of LoA should be handcrafted for that particular milieu. A modicum of middle ground consists of taking a prior LoA, even one that might already be grounded in a specific domain, and mindfully adapting the LoA to a new area or as yet unspecified domain.

In short, the varied approaches seem to consist of:

- Seek a generic LoA and apply it to a target domain Y, or
- Take a generic LoA, adjust and reshape it, then fit the result to a target domain Y, or
- Start entirely fresh for a target domain Y and create an LoA from scratch for it, or
- Reuse a domain-specific LoA of X, and transform as warranted to apply to a different domain Y

In theory, whichever path is undertaken does not especially matter, as long as the final result is an appropriate LoA for the target domain of interest. Though that declaration seems perhaps obvious, the concerns oft-expressed are that the starting point can potentially adversely influence the ending point, such that there is a heightened chance of infusing ill-advised or unwelcomed artifacts into an LoA that otherwise via taking an alternative approach would not have been inadvertently enmeshed. This precautionary warning is properly taken for this proposed framework.

In devising a framework of LoA for the law domain and AILR, let's consider this to be the domain Y, a reuse of salient prior efforts has been undertaken for this framework and thusly benefits from lessons gleaned by those prior accomplishments, doing so with a viewpoint of averting being tainted by such precedents. To a great extent, the reuse of a prior domain-specific LoA is extensively relied upon, aptly justified by its widely accepted use and acclaim as a "gold standard" for LoA's, namely the Society for Automotive Engineers (SAE) J3016 *Taxonomy and Definitions for Terms Related to Driving Automation Systems for On-Road Motor Vehicles* [58], which is known globally and accepted worldwide as a standard LoA for Autonomous Vehicles (AVs) and especially self-driving cars [22] [24].

Consider some key facets of the SAE LoA, which posits these six levels of autonomy [58]:
- Level 0: No Driving Automation
- Level 1: Driver Assistance
- Level 2: Partial Driving Automation
- Level 3: Conditional Driving Automation
- Level 4: High Driving Automation
- Level 5: Full Driving Automation

Some notable aspects that will be further addressed in the next section of this paper encompass that the numbering scheme ranges from a low-to-high indication, the numbering starts with zero, there are six designated levels, the naming of the levels is intended to approximately reflect succinctly the nature of the levels, and each level per the details of the standard is considered separate and distinct from the other levels (we will momentarily return to further inspection of this SAE standard).

It is worth noting that core guiding principles were underlying the formulation of the SAE standard, stated as [58]:
- "1. Be descriptive and informative rather than normative.
- 2. Provide functional definitions.
- 3. Be consistent with current industry practice.
- 4. Be consistent with prior art to the extent practicable.
- 5. Be useful across disciplines, including engineering, law, media, public discourse.
- 6. Be clear and cogent and, as such, it should avoid or define ambiguous terms."

Note that the fifth guiding principle mentions that the SAE standard was intended to be used across disciplines, including the law domain. As such, this framework explicitly lev-erages the SAE standard and does so with appreciation that those having formulated the SAE standard had the forethought to anticipate the added value of their LoA being reused accordingly, avoiding the need to perhaps reinvent the wheel, as it were.

As an example of an effort at reusing the SAE LoA, consider this indication by Yang et al [69] in the context of creating an LoA for the domain of medical robots: "The regulatory, ethical, and legal barriers imposed on medical robots necessitate careful consideration of different levels of autonomy, as well as the context for use. For autonomous vehicles, levels of automation for on-road vehicles are defined, yet no such definitions exist for medical robots. To stimulate discussions, we propose six levels of autonomy for medical robotics as one possible framework."

Here are the six levels that were postulated [69]:

- "Level 0: No Autonomy
    This level includes tele-operated robots or prosthetic devices that respond to and follow the user's command.
- Level 1: Robot Assistance
    The robot provides some mechanical guidance or assistance during a task while the human has continuous control of the system
- Level 2: Task Autonomy
    The robot is autonomous for specific tasks initiated by a human.
- Level 3: Conditional Autonomy
    A system generates task strategies but relies on the human to select from among different strategies or to approve an autonomously selected strategy.
- Level 4: High Autonomy
    The robot can make medical decisions but under the supervision of a qualified doctor
- Level 5: Full Autonomy
    This is a "robotic surgeon" that can perform an entire surgery."

Yet another example consists of efforts by Galdon et al [30] in the use case of Virtual Assistants:
Level 1: No Autonomy
Level 2: Assistance
Level 3: Partial Autonomy
Level 4: Conditional Autonomy
Level 5: Relational Autonomy
Level 6: High Autonomy
Level 7: Full Autonomy

Parasuraman et al [53] had before the SAE standard sought to indicate a generic LoA, consisting of these ten levels:
1. The computer offers no assistance, human must take all decision and actions
2. The computer offers a complete set of decision/action alternatives
3. Narrows the selection down to a few
4. Suggests one alternative
5. Executes that suggestion if the human approves
6. Allows the human a restricted time to veto before automatic execution
7. Exercise automatically, then necessarily informs the human
8. Informs the human only if asked
9. Informs the human only if it, the computer, decides to
10. The computer decides everything, acts autonomously, ignoring the human

As generally might be evident, by-and-large, the number of autonomous levels is usually in the five to ten range, and the preponderance of the approaches conforms to a low-to-high convention.

An important feature of the SAE standard that might not be immediately apparent is the concept of an Operational Design Domain (ODD). An ODD is defined by the SAE standard as this [58]: "Operating conditions under which a given driving automation system or feature thereof is specifically designed to function, including, but not limited to, environmental, geographical, and time-of-day restrictions, and/or the requisite presence or absence of certain traffic or roadway characteristics."

The significance of this crucial concept is that it allows for a subdividing of a domain into those portions that might be amenable to autonomous capabilities or that sooner might be amenable. Without such a proviso, it would tend to hamstring a set of levels in an LoA to require that either autonomy is entirely and completely the case at a given level or it is not at all at that level. This kind of take-it-or-leave-it conundrum was a stumbling block to the acceptability of some other LoA's and represented a subtle but vital form of progression in the formulation of an LoA.

The ODD concept will be instrumental into providing a similar benefit for the LoA of this proposed framework, as will be discussed in the next section.

Briefly, here are the SAE standard levels with an indication of their short-form definitional aspects, notably focused on the driving task (referred to as the DDT or Dynamic Driving Task), incorporating an Automated Driving System (ADS), and expected to provide an OEDR (Object and Event Detection and Response) [58]:

- Level 0: No Driving Automation
    "The performance by the driver of the entire DDT, even when enhanced by active safety systems."

- Level 1: Driver Assistance
    "The sustained and ODD-specific execution by a driving automation system of either the lateral or the longitudinal vehicle motion control subtask of the DDT (but not both simultaneously) with the expectation that the driver performs the remainder of the DDT."

- Level 2: Partial Driving Automation
    "The sustained and ODD-specific execution by a driving automation system of both the lateral and longitudinal vehicle motion control subtasks of the DDT with the expectation that the driver completes the OEDR subtask and supervises the driving automation system."

- Level 3: Conditional Driving Automation
    "The sustained and ODD-specific performance by an ADS of the entire DDT with the expectation that the DDT fallback-ready user is receptive to ADS-issued requests to intervene, as well as to DDT performance-relevant system failures in other vehicle systems, and will respond appropriately."

- Level 4: High Driving Automation
    "The sustained and ODD-specific performance by an ADS of the entire DDT and DDT fallback, without any expectation that a user will respond to a request to intervene."

- Level 5: Full Driving Automation
    "The sustained and unconditional (i.e., not ODD-specific) performance by an ADS of the entire DDT and DDT fallback without any expectation that a user will respond to a request to intervene."

For further details about the SAE standard, including its various limitations and weaknesses, see the in-depth analysis of Eliot [22]. Salient aspects of the SAE standard, considered a specific-domain LoA, will be reused and transformed for purposes of devising the LoA for AILR, as will be indicated in the next sections.

One additional aspect to be covered briefly, particularly when discussing an LoA for the law, entails whether it might be feasible to reuse an existent accepted overarching ontology of the law. Thus, just as reusing an LoA offers merits, so too would reusing an overarching ontology of the law. For clarification, the meaning of ontology in this context is as per Neches et al [52]: "An ontology defines the basic terms and relations comprising the vocabulary of a topic area as well as the rules for combining terms and relations to define extensions to the vocabulary."

As legal scholars are aware, there is not a single unified ontology of the law, though many efforts have been undertaken to form such a taxonomy. For a detailed explanation of ontologies associated with the law, including their strengths and limitations, see [26] [56] [57] [65].

## 3. Foundation for LoA Framework Robustness

In this section, a discussion about the key characteristics that are advisably used when creating a set of autonomy levels is undertaken and includes an examination of how those factors are salient to be used in devising a set of levels of autonomy in the matter of AI Legal Reasoning.

### 3.1 Key Characteristics

When defining levels of autonomy, there is a multitude of factors that should be employed, doing so to systematically arrive at a parsimonious set that is logically sound and inherently robust. Any notable facets that are omitted or skirted, whether inadvertently or by intent, could undermine the veracity of the definition and thus weaken or entirely vacate the utility of the resulting taxonomy.

Utilized here is a bounded set of ten specific characteristics that are significant overall, and for which have been contributory in deriving the levels of autonomy for AI Legal Reasoning. Note that each such characteristic is valuable on its own merits and the listing of them in a numbered or sequenced fashion is not done to showcase priority or ranking, and instead merely showcased for ease of reference.

Those ten key characteristics are:
1. Scope
2. Sufficiency of Reason
3. Completeness
4. Applicability
5. Usefulness
6. Understandability
7. Foolproofness
8. Observe Occam's Razor
9. Differentiable
10. Logical Progression

### 3.1.1 Scope

Scope is a crucial factor since the nature of the underlying act or tasks that are being subject to autonomous operation must be relatively well-stated and apparent to those that seek to rely upon or apply a framework embodying levels of autonomy.

Here, the scope consists of all forms of legal reasoning. This is readily stated but certainly less amenable to being entirely articulated.

Some have argued for example that legal reasoning is essentially that which lawyers do, and therefore the presumed scope would be those acts or effort for which attorneys undertake [9] [10]. But this raises the question of whether an attorney that is say calculating the number of billable hours on a legal case is performing a legal reasoning task, which on the surface does not seem so, and yet falls within the broad interpretation of suggesting that legal reasoning is scoped as that which lawyers are apt to perform.

One remedy would seem to be the addition of a qualifier that legal reasoning is that which lawyers do *when it comes to the practice of law*. This aspect is illuminated via the ABA model definition of the practice of law [68]: "The 'practice of law' is the application of legal principles and judgment with regard to the circumstances or objectives of a person that require the knowledge and skill of a person trained in the law."

Though seemingly somewhat self-referential and thus a bit unclear, the ABA model is further clarified by its attempt to specify the acts or tasks involved in the practice of law, consisting of these four stipulations [68]: "(1) Giving advice or counsel to persons as to their legal rights or responsibilities or to those of others; (2) Selecting, drafting, or completing legal documents or agreements that affect the legal rights of person; (3) Representing a person before an adjudicative body, including, but not limited to, preparing or filing documents or conducting discovery; or (4) Negotiating legal rights or responsibilities on behalf of a person."

For the moment, assume that this provides a general semblance of the scope as it applies to the legal reasoning undertaken by those that formally practice the law as attorneys.

There is still the matter of legal reasoning as utilized by others, including for example judges, which herein is assumed to also be within the scope of these levels of autonomy.

Plus, there is the notion of legal reasoning as used by juries.

Some argue fervently that jurors are not an instance of bona fide legal reasoning per se, apparently being something else instead, perhaps exercising solely common-sense reasoning and not considered equated to the domain-specific elements of legal reasoning [63]. Nonetheless, one can easily argue that there is some form of legal reasoning being relied upon by jurors, regardless of any lack of training in the law or being certified in the law, and therefore it would seem fallacious to excise jurors as legal reasoners altogether.

This discussion raises too the fluency and fluidity properties underlying legal reasoning. A juror might not be entirely fluent in the law and can only muster say a small percentage of their reasoning as being within the realm of legal reasoning when acting in their juror capacity. Does a minimal composition of the harking of legal reasoning somehow play into whether legal reasoning is being deployed, or will any amount, even if infinitesimal, be considered as substantive to being encompassed within the legal reasoning captive?

Such arduous and contentious questions are covered in other versed discussions [46] [50] [61] [66] and are too lengthy to try and settle herein, thus let's proceed to stipulate that any smattering of legal reasoning, regardless of in-the-small or in-the-large, ultimately is considered within the confines of the levels of autonomy for legal reasoning for purposes of the framework being outlined in this paper.

### 3.1.2 Sufficiency of Reason

Sufficiency of reason is a factor entailing whether the levels of autonomy can abide by the Leibniz-like notion of modus ponens inference, generally meaning that each of the asserted levels must have a sufficient explanation for why it is said to be needed or occur.

Any level that does not have adequate justification or rationale would seem unneeded and therefore has no rightful place in the set of levels.

### 3.1.3 Completeness

Completeness is a factor that necessitates assuring that the levels of autonomy can provide a totality of coverage over the realm being subsumed.

If the set of levels does not adequately encompass the scope, this means that the autonomy description is unable to express a fullness of coverage and will suffer accordingly, i.e., by leaving out portions or failing to cope with all that which needs to be specified. The autonomy levels need to embody the entirety of the scope and not have any omitted or overlooked aspects.

That being said, it is equally crucial and certainly preferential to not overshoot the scope and thus by design or by an ill-devised scheme draw into the levels of autonomy those matters that are not in the realm of that which is at hand.

### 3.1.4 Applicability

Applicability refers to an assurance that the levels of autonomy are applicable or practical in their application.

If a set of levels of autonomy are exclusively abstract and unable to be applied, they would seem less valuable than otherwise might be the case. Though such a set might be handy for scholarly pursuits and conceptual analyses, it is argued here that the levels of autonomy have to also be seen as and must readily be able to be applied to that which is considered usable in the real world.

In this case, herein, the levels of autonomy need to be applicable to the day-and-day matters of legal reasoning and be similarly applicable to the broader acts of conceptualizing legal matters too that might arise in academic pursuits regarding the law and legal reasoning.

**3.1.5 Usefulness**

Usefulness is an additive on top of the factor that the levels of autonomy need to be applicable and augments that the set also needs to be useful in its application.

In other words, it might be possible to apply something, and yet it in the end provides little utility in doing so. The consideration here is that the levels must also rise to the occasion and provide usefulness that is part-and-parcel of their existence.

For example, in the case of AI Legal Reasoning, given the existing confusion and confounding state of affairs regarding what vendor offerings provide in the way of computer-based legal reasoning capacities, a set of levels of autonomy could aid in clarifying such matters and therefore serve a quite useful purpose.

Likewise, for those scholars striving to devise advances in AI Legal Reasoning, a set of levels of autonomy that is useful would provide guidance as to where the state-of-the-art presently resides and where the future direction of new efforts can potentially aim.

**3.1.6 Understandability**

Understandability is the nature of how readily comprehended or conveyed the levels of autonomy are.

If the levels are arcane or obtuse, the eventual applicability and usefulness are most likely undermined. In turn, this suggests that the levels of autonomy would not gain awareness and nor take hold as a viable means of defining autonomous operations.

That being said, some might argue that there is not a need for a set of autonomy levels to be popular and that undue admiration toward seeking popularity might water down or subvert a rigorous approach. In one sense, those kinds of arguments can be a false portrayal of a misleadingly alluded to mutually exclusive condition. The implication is that rigor can only exist when there is not popularity, while that which is popular cannot somehow include rigor. This presumption needs to be rejected. Instead, the merits of the set of levels of autonomy can be assessed on both its semblance of rigor and its semblance of popularity, both of which can very well co-exist, and perhaps more so if the levels of autonomy are particularly well-designed accordingly (rather than by happenstance).

**3.1.7 Foolproofness**

Foolproofness is a factor that attempts to indicate whether the levels of autonomy can be too readily distorted or twisted to accommodate those that might wish to subvert the set, or in the counter, whether the levels are strongly devised to reduce the ease of subversions.

This is a fundamental recognition that there will be those that wish to misuse the levels for purposes of making claims beyond which they should not be allowed to do. Certainly, such claims are going to be made, no matter how foolproof the structure might be, nonetheless, the notion is to try and anticipate such untoward efforts, and prevent or at least diminish the ease of enabling those saboteurs from doing so.

For example, someone claiming to have an AI legal reasoning system at level Z, while in fact only a level less than Z has been achieved, ought to have hurdles or other barriers that make such a claim readily shown to be false and inappropriate. This capacity can then be handy for revealing those that are making false claims and would undoubtedly be instrumental in aiding those that sincerely are desirous of abiding by the set and not accidentally misjudge the nature of the levels.

**3.1.8 Observe Occam's Razor**

As William of Ockham succinctly stated [67], "plurality should not be posited without necessity," which has become widely known as Occam's razor. In short, everything else being equal, the simpler of two competing approaches ought to weigh toward the favor of that which is the simpler and thus accordingly cast into disfavor that which is not so.

The set of levels of autonomy should be held to the test that it proffers the simplest possible rendering, without loss of other factors, and that anything that is otherwise overcomplicated should be further reduced into simplicity if at all feasible.

As an example, if a set of levels of autonomy were initially devised to be twenty such levels, and if there was a means to reduce the set to say ten, doing so without losing or undermining the set, there would be a preference toward the set of ten in lieu of the larger set of twenty, under the belief that the set of ten is the simpler of the two competing approaches (everything else being equal).

Of course, it is crucial to not blindly seek Occam's razor and neglect to uphold the key premise that everything else being equal overrides the simplicity goal. Sometimes there is an inadvertent race to the bottom, as it were, failing to realize that along the way there has been a substantive loss in other merits of the structure.

**3.1.9 Differentiable**

Differentiable is a factor that involves the clarity of separation or distinction between the levels of autonomy that are being stated.

Suppose there was a level M and a separate level O in a given defined set of levels of autonomy. If there was no ready means to differentiate between level M and level O, this would suggest that there is truly no need for two such levels and they could be consolidated into one level (this abides by the Occam's razor).

Indeed, if the set is allowed to exist with the two separate levels, M and O, yet they are indistinguishable from each other, it would indubitably create confusion as to whether a claimant is rightfully using level M or rightfully using level O, which presumably they could use either one as they so wish, but this then decreases the efficacy of the levels.

Each level ought to stand on its own, separately and distinctly, and not be readily muddled into another level.

**3.1.10 Logical Progression**

Logical progression is perhaps one of the most controversial of these ten factors and entails the supposition that the levels of autonomy should have a preponderance toward a progression or advancement of autonomous capacities.

Suppose that we had a set of autonomous levels that consisted of ten levels. Presumably, the ten levels could be scrambled and placed into any order one might so wish. They essentially could be randomly arrayed.

But this would seem to undercut several of the other factors already mentioned herein, such that if there was a means to sequence or order the levels, it might make the set more readily useful, applicable, etc. This does not imply that a force fit is appropriate, and a false or forced effort to arrange the levels is little better than a random arrangement, one would so argue (perhaps worse so).

It would be preferential to have the levels of autonomy arranged into a logical profession, making the set easier to assimilate and apply.

This might be arrayed from low to high, or from high to low, and does not materially make a difference, though it can be said that prior sets of autonomy have typically gone from low to high, establishing a kind of default approach that is more likely to resonate for any subsequent sets of autonomy.

On an allied aspect, there is also the matter of whether to number the levels or assign them letters of the alphabet or use some other means to designate that the levels are an ordered set and range from low to high or from high to low. Providing an indicator of their respective positioning is abundantly helpful as an aid in referring to the various levels. A preference herein is given to the use of numbering, though one could make use of some other method if so desired.

Part of the handiness of numbering is that we already arithmetically accept and immediately comprehend that numbers illustrate a progression. There is little cognitive effort in making that kind of mental leap and tends to make the set more intelligible and easier to refer to.

When using numbers, the question seems to arise repeatedly about the use of the number zero. In essence, some dislike the use of a zero within the levels of autonomy and suggest it is an artifact of those that are computer-versed that they oft include the number zero (a bits and bytes mindset, as it were), whereas presumably, it is customary that people start usually a counting sequence with the number one, rather than starting at the number zero.

On the other hand, it can be argued that the use of zero is intuitively useful since if it is used to denote that there is no semblance of autonomy, and those that use the set will readily grasp why the set starts with the number zero. This matter about the use of zero might seem like a mundane debate and unworthy of consideration, but do not downplay the significance of how important the numbering can become. By-and-large, most sets of autonomy become known by their numbering scheme, even more so than any naming or descriptor that is associated with each of the levels.

Another seemingly debated topic is whether the levels should have numbers-only or whether they should have names or descriptors only. Again, this seems to be a false dichotomy. Nothing is preventing the use of both numbers and a name or descriptor, and indeed this is generally customary as an approach, and suitable too as it provides the convenience and shortcut use of a number to denote a level, along with allowing for a name or descriptor that can provide more substance.

The naming or descriptor has to also be carefully worded and be mindfully expressed. If the words used to name or describe a level are inexpressive, they can undermine the entire set and also create confusion over the significance of the numbering scheme being used. Thus, the wording should succinctly designate the overarching significance of the level, using as few words as possible, and the right words, and not use overloaded words or be superfluous in the wording.

One additional concern has to do with the logical progression and whether the stepwise movement from one level to the next is somehow a simple linear movement. The problem with numbers is that we typically think of the number 2 as simply one more than the number 1, and the number 3 to be simply two more than the number 1, and so on. This causes difficulty in expressing for example the magnitude of earthquakes, which the famous Richter scale attempts to do, and for which the numbers though seemingly progressing one at a time represent a much larger magnitude in jumps.

Some would assert that the numbering ought to convey magnitudes of degree or change when it is so built-in to the levels, and therefore that perhaps instead of numbering 1, 2, 3, 4, and so on, the numbering should be something like 1, 20, 100, etc., attempting to immediately illustrate the magnitude differences. Unfortunately, this attempt at overcoming one facet then introduces other problems, such as the difficulty of people remembering what the numbers of the levels are and tends to undermine their use accordingly. As such, the convention seems to be that the use of ordinary counting numbers is easiest to be conveyed and be remembered, and meanwhile, there should be an ongoing effort to try and communicate that the levels are of increasing orders of magnitude.

**4. Autonomous Levels of AI Legal Reasoning**

In this section, a proposed framework for the autonomous levels of AI legal reasoning is depicted. In addition to the depiction, there is also an indication of how the devised autonomous levels conform to the key characteristics discussed in the prior section, providing a rationale for understanding the basis of the formulated levels and what they portend to provide.

The proposed levels are as follows:

Level 0: No Automation for AI Legal Reasoning

Level 1: Simple Assistance Automation for AI Legal Reasoning

Level 2: Advanced Assistance Automation for AI Legal Reasoning

Level 3: Semi-Autonomous Automation for AI Legal Reasoning

Level 4: Domain Autonomous for AI Legal Reasoning

Level 5: Fully Autonomous for AI Legal Reasoning

Level 6: Superhuman Autonomous for AI Legal Reasoning

**4.1 Background and Rationale**

A fundamental question involves how many levels ought to be used for adequately depicting the levels of autonomy of AI Legal Reasoning.

We can begin by first stating the seemingly obvious, namely that there would certainly seem to be at least two such levels, namely a level of which there is no legal reasoning and a second level in which there is in fact some amount of legal reasoning. In that sense, the minimum set would be at two levels.

It is possible to try and argue that there only needs to be one level and that the absence of autonomy for legal reasoning is implied, but this seems somewhat disingenuous as an assertion and preferably better handled explicitly rather than by implicit default or base assumption.

Beyond the core of the two foundational levels, the second level consisting of the legal reasoning of some amount could readily be argued as worthy of further subdivision. Furthermore, if the full set is to be considered complete, it would seem logical to suggest that there should be a topmost level. In that way of thinking, the levels now should be a set of three, consisting of a level of no legal reasoning, a level of some amount of legal reasoning, and a third or uppermost level of full legal reasoning or a pinnacle level.

This then arrives at a minimum set of three levels. Are there are more levels needed? Abiding by the earlier desire to observe Occam's razor, any additional levels would need to be carefully and thoughtfully proffered, assuming that three levels alone might be accommodating to the matter. Simplicity is the watchword.

One especially salient aspect that has arisen in the AI field is whether there will be the chance that AI will surpass human intelligence and proceed into some form of superhuman intelligence level [23] [28]. Though this is certainly debatable, it would seem prudent to prepare the levels of autonomy for such a possibility, regardless of whether it might be viable in the near-term or not, and thus be prepared for the long-term in case the superhuman facets materialize.

In that case, it would be prudent to add a fourth level, a superhuman level of autonomy to the set of AI legal reasoning levels. As might be evident, this now suggests that the minimum set of levels is four.

One quick aside is that some might have qualms over this kind of logical bottoms-up construction of the levels and want to declare summarily that the levels are the levels, meaning that a top-down approach should be utilized instead. This presumably implies that we can merely inspect the legal reasoning realm and will somehow naturally divine the appropriate number of levels. It will accountably appear by the act of inspection alone.

This top-down perspective as a starting point is worthy of consideration, though if the bottoms-up method arrives at ultimately the same set, it would seem to not make any substantive of a difference as to how the set was arrived at. Whichever approach is undertaken, the other approach would be a helpful double-check. If one wants to try and make the case that one approach is more intrinsically advantageous or quicker to the answer being sought, that's fine, though it does not materially impact the result and indeed should have no effect whatsoever.

With four levels now in hand, once again the question arises as to whether this is sufficient, complete, understandable, applicable, and so on. Let's consider these facets.

One of the greatest downsides of any set of levels of autonomy is the confluence of everyday *automation* with the kind of automation intended via the use of AI. The very act of saying that there is a level of autonomy does tend to imply that *autonomous* effort is present, and regrettably introduces confusion and the potential for untoward uses of the set.

It is helpful for the factors of at least understandability and applicability to try and separate the levels containing autonomy from those that do not. In that sense, we now need five levels, namely one with no automation, one with everyday automation, one with autonomy at some level, one with full autonomy equivalent to human intelligence, and one at a superhuman level.

This aids in abiding by the factors, but then opens the question about the possible subdivisions within the notion of everyday automation. As advances are made in non-AI automation, it would seem unruly to cast all such automation as essentially being the same. As such, it would be prudent to add a level to allow for stipulating a simple variant of everyday automation and an additional level for more advanced automation.

For those levels that are lacking in AI autonomous capacities, the assumption is that the automation will be assistive rather than performing autonomously, which makes definitional sense herein. In that vein, the naming or descriptor for those levels should be careful in not conflating the notion of *automation* versus the notion of *autonomy*.

In the case of full autonomy, the requirement of AI to achieve full autonomy is undeniably aspirational in this context. To aid in the levels providing coverage for that which is less than full autonomy, we will add another level. This additional level will allow for domain-specific AI legal reasoning instantiations, similar to the earlier discussed ODDs in the case of AV's.

Finally, there is the situation of AI that is semi-autonomous. Some would argue, at times persuasively, that there should not be a level or category known as semi-autonomous. The argument goes that this is a slippery middle ground and lends itself to being misused. A viewpoint taken is that something is either autonomous or it is not and trying to split hairs by including a semi-autonomous grouping is perilous and unsettling.

Those are valued words of caution. Nonetheless, in a practical sense, there is a gray area into which there is some amount of AI that goes beyond ordinary automation, and then there is more robust AI that takes the realm into the autonomous sphere. Somehow, the gray area needs to be included and not omitted.

Thus, though freely acknowledging the potential drawbacks, it seems more so beneficial to include a semi-autonomous category than to exclude it.

This brings up an allied topic which is that there will inexorably over time be a shift of what we all accept as ordinary automation versus that which is considered AI. In that way, any set of levels of autonomy might very well need to be refined and adjusted, though that will play out over a lengthy period and does not negate or undermine the value of the levels at a point in time.

All told, based on the foregoing, these again are the proposed levels:

Level 0: No Automation for AI Legal Reasoning

Level 1: Simple Assistance Automation for AI Legal Reasoning

Level 2: Advanced Assistance Automation for AI Legal Reasoning

Level 3: Semi-Autonomous Automation for AI Legal Reasoning

Level 4: Domain Autonomous for AI Legal Reasoning

Level 5: Fully Autonomous for AI Legal Reasoning

Level 6: Superhuman Autonomous for AI Legal Reasoning

**4.2 Explanation of the Levels**

There are seven levels in the proposed framework.

In the matter of whether this is possibly excessive and a lesser number of levels might be more readily grasped, the aspect that the levels are at the count of seven is in conformance with the well-known so-called magical number 7 and plus-or-minus 2, a longstanding classic rule-of-thumb established in research on human psychology by Miller [48]. Also, each of the levels is justifiable on a standalone basis and the levels are progressively arranged in a logical order from low to high in terms of autonomous capabilities, thus proffering a relatively easily understood structure.

It could be argued that there is a somewhat natural flow and coherence to the devised levels and that it abides by the "Goldilocks principle" of just the right number of levels, none less so and none more so.

As an aside, it is acknowledged that one minor problem associated with starting the autonomous levels numbering at the number zero is that it inadvertently creates some potential confusion over how many levels there are. With similar such numbering, it is easy for many to simply look at the highest number of the levels, in this case, the number six, and assume therefore that there are only six levels. This is a recurring aggravation and source of some mild disorientation for other instances of the use of zero as a starting point. In any case, it can be argued that the use of level zero is still of merit and those that misquote or misstate the number of levels are doing so by lack of awareness, plus it does not necessarily hamper or undermine the set, other than at times sparking modest confusion about how many levels there are in the set.

In terms of the levels, Level 0 is the special case of no automation, meaning that there is no notable automation involved in undertaking AI and legal reasoning.

The basis for stating that there is no notable automation stems from a debate over whether say a fax machine used in transmitting legal documents fits into Level 0 or would fit into Level 1. Strictly speaking, since a fax machine could be envisioned as a form of automation, it would seemingly belong in Level 1, though others contend that a fax machine does not rise to the notion of automation and should, therefore, be cast into Level 0. This overarching question about the boundaries and assessing fit for clarity is addressed in the respective subsections.

Level 1 and Level 2 entail automation of an ordinary manner that would be hard-pressed to be described as AI capabilities, while Level 3 is the in-between state of automation that is approaching AI-like capacities, yet still not within the realm of autonomy.

Level 4 is the first of the designated autonomous levels and consists of allowed-for constraints upon the range or scope of the autonomy, indicating that within a specified subdomain of law the AI legal reasoning can operate autonomously. Level 5 is considered full autonomy in terms of AI legal reasoning across all domains of the law and encompassing an entirely autonomous operation. Level 6 is the superhuman autonomous level of the law as undertaken by AI legal reasoning, and for which it is unknown if any such autonomous capability will ever be achieved but has nonetheless been included for completeness' sake.

**See Figure 1 and Figure 2 for a summary chart depicting the autonomous levels of AI Legal Reasoning**.

**Figure 1** indicates via rows of the chart the successive levels of the framework and then depicts the main name or descriptor, followed by an exemplar short set, and then a brief indication of the automation capacity that is then followed by the latest status. The latest status column will naturally change over time in terms of its contents (due to advancements in technology and usage), while the other columns will remain static and are deemed definitional and unchanging thereof.

**Figure 2** is similar to Figure 1, though showcases the same material via placing the levels upon the columns. This alternative portrayal is intended merely to help present the same information in a different format, a convenience of presentation or display, and not meant to introduce any new or differing facets or content. In that manner, the two figures are wholly consistent and aligned with each other.

### 4.2.1 Level 0: No Automation for AI Legal Reasoning

Level 0 is considered the no automation level. Legal reasoning is carried out via manual methods and principally occurs via paper-based methods.

This level is allowed some leeway in that the use of say a simple handheld calculator or perhaps the use of a fax machine could be allowed or included within this Level 0, though strictly speaking it could be said that any form whatsoever of automation is to be excluded from this level.

If purity of exclusion helps to avoid attempts at misusing the Level 0, it would seem prudent to take such a stark position, though there is a balance required between being dogmatic and yet allowing for some flexibility in the spirit and denotation of the levels. It seems doubtful though, in any case, that many would seek to argue about Level 0 versus Level 1, since those levels are rather straightforward and without particular acclaim, and thus the need to bear down on being strictly stipulated would not seem especially bothersome or significant to be entertained for these ascertained levels.

On a related aspect, keep in mind that the levels can apply to different facets of the act of practicing law. For example, within a law office, there might be some tasks done entirely by manual and paper-based methods, while other tasks being carried via presumed forms of automation. Some would be falsely quick to ascribe that the law office is to be rated as at a Level 0 of the levels of autonomy since there are some instances of no use of automation. Others might insist that since there is some amount of presumed bona fide automation in use, the entirety of the law office should be given a Level 1 rather than a Level 0 designation.

This is an unfortunate misreading and misinterpretation of the intentions of the levels of autonomy. It would be expected that those tasks of the law office without the use of automation are considered at a Level 0 and meanwhile, simultaneously, the other tasks of the law office using automation might very well be at Level 1 or perhaps Level 2, or higher. There is nothing inconsistent or incoherent about the application of the levels as being applied to particular segments or portions of activity.

In that same vein, clarification is perhaps for the viewpoint of that which is an instance versus that which is a generalized facet. For example, suppose a law office is performing a task via manual methods, doing so by their choice to do so, and yet suppose further that there is automation that could be applied to those tasks, but the law office has not yet opted to adopt such automation. It would be a mischaracterization to then say that those tasks are Level 0 per se since there is in fact (we are assuming) automation available that could carry out or assist in those tasks.

As such, it would be preferred that the use of the autonomous levels be used in a generalized fashion, demarking the state of the whole, rather than being used in the assessment of a particular practice or usage.

**4.2.2 Level 1: Simple Assistance Automation for AI Legal Reasoning**

Level 1 consists of simple assistance automation for AI legal reasoning.

Examples of this category encompassing simple automation would include the use of everyday computer-based word processing, the use of everyday computer-based spreadsheets, the access to online legal documents that are stored and retrieved electronically, and so on.

By-and-large, today's use of computers for legal activities is predominantly within Level 1. It is assumed and expected that over time, the pervasiveness of automation will continue to deepen and widen, and eventually lead to legal activities being supported and within Level 2, rather than Level 1.

The demarcation between Level 1 and Level 0 has been discussed in the Level 0 subsection, while the demarcation between Level 1 and Level 2 is discussed next in the Level 2 subsection.

**4.2.3 Level 2: Advanced Assistance Automation for AI Legal Reasoning**

Level 2 consists of advanced assistance automation for AI legal reasoning.

Examples of this notion encompassing advanced automation would include the use of query-style Natural Language Processing (NLP), Machine Learning (ML) for case predictions, and so on.

Gradually, over time, it is expected that computer-based systems for legal activities will increasingly make use of advanced automation. Law industry technology that was once at a Level 1 will likely be refined, upgraded, or expanded to include advanced capabilities, and thus be reclassified into Level 2.

The demarcation between Level 1 and Level 2 is undoubtedly likely to spur great debate and consternation. Vendors of legal technology are more desirous of having their wares classified as Level 2 versus at Level 1. To try and prevent or head-off this difficulty, it would certainly be preferable to have an ironclad set of metrics or stipulations that would rule out that which is attempting to misleadingly attempt to be labeled as Level 2 when it is more reasonably stated as Level 1.

One approach to coping with this dilemma would be to enumerate all possible kinds of legal technology that falls within Level 1 and within that of Level 2, thus, it would be a simple matter of ensuring that a given legal technology either matched to those listed in the Level 1 definition or matched to those in the Level 2 definition.

This same debate arises in trying to discern Level 2 versus Level 3, and therefore is a recurring problematic consideration that permeates not only this set of autonomous levels but generally occurs in any set of autonomous levels. In essence, autonomous levels tend to defy any simple indication of metrics or enumeration that could delineate indisputable crafting of scope and boundaries that manifestly distinguishes one level versus another.

For the moment, we will lean into the use of a reasonableness test, namely that some semblance of reasonableness concerning the overall spirit and intent of the levels of autonomy is to be observed. This is an open research question too as to how it can be ultimately finitely resolved, if it can, and might consist of some anointed standards bodies that ascertain the specifics of what constitutes each level and rates or judges submissions of legal technology that is seeking a form of certification for their claimed achieved level. For those seeking a more precise and perhaps mathematical or formulaic distinction, this topic is certainly a worthwhile research pursuit to determine if such an approach is potentially viable and workable.

**4.2.4 Level 3: Semi-Autonomous Automation for AI Legal Reasoning**

Level 3 consists of semi-autonomous automation for AI legal reasoning.

Examples of this notion encompassing semi-autonomous automation would include the use of Knowledge-Based Systems (KBS) for legal reasoning, the use of Machine Learn-ing and Deep Learning (ML/DL) for legal reasoning, and so on.

Today, such automation tends to exist in research efforts or prototypes and pilot systems, along with some commercial legal technology that has been infusing these capabilities too.

All told, there is increasing effort to add such capabilities into legal technology and thus it is anticipated that many of today's Level 2 will inevitably be refined or expanded to then be classifiable into Level 3.

The same debate about what belongs in Level 2 versus Level 3 is akin to the debate about what belongs in Level 1 versus Level 2 and has been covered ergo in the discussion about Level 2 (see prior subsection). Once again, the answer, in brief, is that there is a reasonableness test to be assumed and that for now, there is no formulaic or precise demarcation, subject to further research and consideration.

### 4.2.5 Level 4: Domain Autonomous for AI Legal Reasoning

Level 4 consists of domain autonomous computer-based systems for AI legal reasoning.

This level reuses the conceptual notion of Operational Design Domains (ODDs) as utilized in the autonomous vehicles and self-driving cars levels of autonomy, though in this use case it is being applied to the legal domain.

Essentially, this entails any AI legal reasoning capacities that can operate autonomously, entirely so, but that is only able to do so in some limited or constrained legal domain.

Some elaboration on these aspects might help ensure that Level 4 is well understood.

First, unfortunately, there is no globally accepted standardized way to stipulate what the legal domains per se consist of.

Efforts to ontologically specify law have been made repeatedly, and there are many approaches to choose from, but there does not seem to be one wholly accepted and nor fully adopted taxonomy that could be leveraged for this Level 4 definition, particularly without lively debate and inexorably falling into a related but not integral definitional abyss that would tend to undermine the overarching aspirations of this framework, needlessly so.

As an example of what types of legal domains might be construed, consider these [27]:
- Animal law
- Admiralty law
- Bankruptcy law
- Banking law
- Civil Rights law
- Constitutional law
- Corporate law
- Criminal law
- Education law
- Entertainment law
- Employment law
- Environmental law
- Family law
- Health law
- Immigration law
- International law
- IP law
- Military law
- Personal injury law
- Real Estate law
- Tax law
- Etc.

The indicated list of potential legal domains is not exhaustive and could readily be further expanded and refined. Another perspective on legal domains could be by delineating the type of task performed, such as these functional areas of law practices [27]:
- Case Management
- Contracts
- Courts/Trials
- Discovery
- Documents/Records
- IP
- Law Office/Practice
- Lawyer & Client Interaction
- Legal Assistants
- Legal Collaboration
- Legal Research
- Legal Workflow
- Legal Writing
- Professional Conduct

There is also the matter of what is a domain in terms of its degree of magnitude. For example, would Case Management for Real Estate law to be considered a domain or a subdomain? This raises the question about the extent of domains and also the extent of subdomains, along with the perhaps ad infinitum possibility of subdomains within subdomains, etc. Envision a subdomain of a subdomain of a subdomain of a subdomain that is so narrow in scope that it would perhaps be easy or nearly trivial to claim that there is autonomous AI legal reasoning that someone has crafted for that tiny milieu.

In brief, this is an open question as to what the domains or subdomains would consist of, and thus further research is desired and necessary to aid in pinning down the particulars for Level 4. That being said, it is worth noting that the autonomous vehicles and self-driving cars ODD is similarly without definitive stipulation about the parameters of the domains, and yet the autonomous levels and Level 4 are generally considered agreed to and put into use. The point is that even if part of a framework is left open, for now, this does not negate the framework in any demonstrative way and instead simply leaves available the utility of closing the gap by rendering some later specificity to the matter.

#### 4.2.6 Level 5: Fully Autonomous for AI Legal Reasoning

Level 5 consists of fully autonomous computer-based systems for AI legal reasoning.

In a sense, Level 5 is the superset of Level 4 in terms of encompassing all possible domains as per however so defined ultimately for Level 4. The only constraint, as it were, consists of the facet that the Level 4 and Level 5 are concerning human intelligence and the capacities thereof. This is an important emphasis due to attempting to distinguish Level 5 from Level 6 (as will be discussed in the next subsection)

It is conceivable that someday there might be a fully autonomous AI legal reasoning capability, one that encompasses all of the law in all foreseeable ways, though this is quite a tall order and remains quite aspirational without a clear cut path of how this might one day be achieved. Nonetheless, it seems to be within the extended realm of possibilities, which is worthwhile to mention in relative terms to Level 6.

#### 4.2.7 Level 6: Superhuman Autonomous for AI Legal Reasoning

Level 6 consists of superhuman autonomous computer-based systems for AI legal reasoning.

In a sense, Level 6 is the entirety of Level 5 and adds something beyond that in a manner that is currently ill-defined and perhaps (some would argue) as yet unknowable. The notion is that AI might ultimately exceed human intelligence, rising to become superhuman, and if so, we do not yet have any viable indication of what that superhuman intelligence consists of and nor what kind of thinking it would somehow be able to undertake.

Whether a Level 6 is ever attainable is reliant upon whether superhuman AI is ever attainable, and thus, at this time, this stands as a placeholder for that which might never occur. In any case, having such a placeholder provides a semblance of completeness, doing so without necessarily legitimatizing that superhuman AI is going to be achieved or not. No such claim or dispute is undertaken within this framework.

### 4.3 Magnitudes of the Levels

As earlier stated, there is always a complication that using a numbering scheme of simple integers for a set of autonomous levels can convey an implied equal magnitude difference between the levels. It is overly easy for someone to construe that say Level 2 is merely one more than Level 3, thus the leap or jump is of some unstated magnitude, and that likewise the movement from say Level 3 to Level 4 is the equal amount of a shift or step-up increment when perhaps the chasm between those respective levels is uneven and dramatically differs. Trying to use the numbering scheme to suggest magnitudes is unfortunately overloading that then tends to undermine the simplicity and ease of conveying what the levels are.

The numbering for this framework falls within that same approach of using simple incremental integers, and yet the magnitude between the levels is uneven per each jump from level to level.

As illustrative of the span between the steps see Figure 3.

**Figure 3** is not drawn to scale and merely anecdotally is presented to suggest that there are magnitudes of difference between each step. The curve shown could be redrawn in a multitude of ways, and there is an argument to be made that Level 6 might be so far off the chart that you would need to shrink the rest of the graph into a tiny smidgen to get Level 6 onto the chart at all.

In any case, the point is that it is vital to realize that there are varying magnitudes of difference between each of the levels and it is not a simple linear progression among them.

### 4.4 Conformance to Key Characteristics of Autonomy Levels

Recall that the earlier portion of this section proffered these suggested key characteristics that a sound and robust set of autonomy levels should aspire to attain:

1. Scope
2. Sufficiency of Reason
3. Completeness
4. Applicability
5. Usefulness
6. Understandability
7. Foolproofness
8. Observe Occam's Razor
9. Differentiable
10. Logical Progression

In a recap of the seven proposed autonomous levels for AI Legal Reasoning, it is hoped that the preceding discussion about the levels is adequate to showcase conformance to those key characteristics. The scope was directly discussed, as were the reasons for each level and the entire set, plus a semblance of completeness, applicability, usefulness, understandability, foolproofness, application of Occam's razor, differentiability, and logical progression were indicated.

That being the case, and for clarification, meeting those characteristics to whatever degree has been achieved does not necessarily mean that for all intent and purposes that a framework is settled or somehow conclusively stated. That is decidedly not the case herein, and much additional work and open questions are still to be addressed, but this is to be realized and does not impinge or undermine the essence of the framework and nor hamper or mar its initial introduction and formulation.

## 5. Additional Considerations and Future Research

In this final section, coverage of additional facets is included and so too are some strident calls for future research for the furtherance of this important topic.

First, note that the framework is considered descriptive rather than prescriptive.

Second, nothing about the levels and nor the framework is intended to suggest that legal reasoning autonomy will indeed be achieved via AI. Some are predicting that a so-called *legal singularity* will someday arise, purportedly denoting a time at which the laws are entirely established and adjudicated via AI autonomous systems. Within those predictions, there is a concern that the law might inevitably become cast in stone, unchanging and unwavering, or that the law will be so stipulated and codified that all legal uncertainty is excised and thus the entirety of legal outcomes is perfectly predictable.

This framework is neither supportive of such assertions and nor a denier of such theories. In a sense, the framework is the framework, intending simply to provide a means of identifying and distinguishing levels of autonomy concerning the law and AI legal reasoning. Whether such autonomy is achieved, or when it might be achieved, does not bear on the nature of the framework. For example, it could very well be that no AI system ever is sufficiently capable to be considered at a Level 5, but that does not negate the value of having a Level 5 as part of the framework.

Likewise, for those that especially eschew the concept of a superhuman AI capability, having included this element as a cornerstone of Level 6 is not somehow a testament that superhuman AI will be reached. Construe Level 6 as a future placeholder, potentially sitting empty for a time to come, yet nonetheless available if the day should ever arrive for its use.

On a related theme, there is much discussion in and beyond the law industry concerning whether lawyers and legal professionals will ultimately be replaced by AI autonomous legal reasoning. This kind of societal consideration is again another aspect that is not within the purview of this framework. No commentary or weighing in about the matter is substantive to the rendering and applicability of this LoA. That's not to suggest that such a topic is not erstwhile, only that it is not pertinent to the formulation of this specific matter at hand.

In terms of scope of this LoA, those in the field of law are at times focused on what is somewhat euphemistically referred to as the shadow of the law [49], and as such, there might be a question as to whether this particular dimension of the law is also within the scope of the framework. In short, yes, the shadow of the law would also be encompassed to the degree that it involves the crucial ingredient that underlies all facets of the intended scope, namely the instantiation of legal reasoning.

For each of the levels, an attempt has been made to showcase that each level is distinct from the other levels. One of the frequently raised questions about any LoA is whether it is possible to be a partial member of a given level, sometimes denoted by a fractional amount. For example, perhaps a legal system software is within Level 2, but the system is nearing to Level 3, thus, the inclination is to proffer that the software is a Level 2.5 or some akin fractional amount. This is decidedly *not* the nature of this framework and typically disallowed indeed by most such frameworks, including the SAE standard for AVs.

Next, consider some typical questions disclosed when evaluating a legal industry LoA.

One concern is whether adopting this kind of framework for the legal industry will expose those using the LoA to a form of *alluring legal liability*.

Suppose that a vendor Q opts to claim that their AI legal system is rated at a Level 4. In terms of potential legal liability, some question whether this opens the vendor to potential legal exposures if it can be later shown that the system was not qualified for a Level 4 rating. In that sense, the preference by some, such as vendors, might be to avoid using the framework, for angst of incurring a legal exposure.

Some are also concerned that a framework such as this LoA might be unduly codified into the law itself, perhaps becoming a regulation that is lawful to be observed. On the one hand, this could be said to provide teeth to any such LoA and aid in promulgating it, but at the same time perhaps serve in an overbearing or stifling manner. Indeed, there are reservations about the potential heralding of a Collingridge dilemma [17] by enacting any such framework. This refers to a postulated theory by researcher Collingridge that suggests if structures and potential burdens are placed onto an innovation before it has a chance to breathe, or maturely innovate and gain traction, this can inadvertently quash or delay the innovation. These are considerations worthy of further research.

Also, a more detailed specification for the nature of each level would be another viable and fruitful avenue of additional research. Doing so is evocative of the likeliest controversy for this kind of LoA, as is similarly the case for most any LoA, which typically comes down to trying to ascertain the veracity of a claim that an AI system has earned its way into a particular level. The tighter that such a measuring mechanism can be devised, presumably it will be easier for those using the framework to accurately and readily select the appropriate level, and also more expediently unmask false designations. As the acclaimed management theorist Peter Drucker has been oft-quoted as asserting [19], you cannot manage that which you cannot measure.

For far too long, the legal industry has been without an accepted and robust Levels of Autonomy (LoA) measuring tool for AILR which this framework proposes a draft formulation of, serving as a potential and earnest step in that needed direction. Additional research is welcomed and highly encouraged in this hoped-for valued contribution to the future of law and aims to provide a potential impetus for or otherwise aid in the advent of automation and autonomy of the law via AI and Legal Reasoning.

**About the Author**

Dr. Lance Eliot is the Chief AI Scientist at Techbrium Inc. and a Stanford Fellow at Stanford University in the CodeX: Center for Legal Informatics. He previously was a professor at the University of Southern California (USC) where he headed a multi-disciplinary and pioneering AI research lab. Dr. Eliot is globally recognized for his expertise in AI and is the author of highly ranked AI books and columns.

# Figure 1

## AI & Law: Levels of Autonomy For AI Legal Reasoning (AILR)

| Level | Descriptor | Examples | Automation | Status |
|---|---|---|---|---|
| 0 | No Automation | Manual, paper-based (no automation) | None | De Facto - In Use |
| 1 | Simple Assistance Automation | Word Processing, XLS, online legal docs, etc. | Legal Assist | Widely In Use |
| 2 | Advanced Assistance Automation | Query-style NLP, ML for case prediction, etc. | Legal Assist | Some In Use |
| 3 | Semi-Autonomous Automation | KBS & ML/DL for legal reasoning & analysis, etc. | Legal Assist | Primarily Prototypes & Research Based |
| 4 | Domain Autonomous | Versed only in a specific legal domain | Legal Advisor (law fluent) | None As Yet |
| 5 | Fully Autonomous | Versatile within and across all legal domains | Legal Advisor (law fluent) | None As Yet |
| 6 | Superhuman Autonomous | Exceeds human-based legal reasoning | Supra Legal Advisor | Indeterminate |

*Figure 1: AI & Law - Autonomous Levels by Rows*     *Source Author: Dr. Lance B. Eliot*     V1.2

**Figure 2**

| | Level 0 | Level 1 | Level 2 | Level 3 | Level 4 | Level 5 | Level 6 |
|---|---|---|---|---|---|---|---|
| | | | | AI & Law: Levels of Autonomy For AI Legal Reasoning (AILR) | | | |
| **Descriptor** | No Automation | Simple Assistance Automation | Advanced Assistance Automation | Semi-Autonomous Automation | Domain Autonomous | Fully Autonomous | Superhuman Autonomous |
| **Examples** | Manual, paper-based (no automation) | Word Processing, XLS, online legal docs, etc. | Query-style NLP, ML for case prediction, etc. | KBS & ML/DL for legal reasoning & analysis, etc. | Versed only in a specific legal domain | Versatile within and across all legal domains | Exceeds human-based legal reasoning |
| **Automation** | None | Legal Assist | Legal Assist | Legal Assist | Legal Advisor (law fluent) | Legal Advisor (law fluent) | Supra Legal Advisor |
| **Status** | De Facto – In Use | Widely In Use | Some In Use | Primarily Prototypes & Research-based | None As Yet | None As Yet | Indeterminate |

*Figure 2: AI & Law - Autonomous Levels by Columns*   *Source Author: Dr. Lance B. Eliot*

V1.2

**Figure 3**

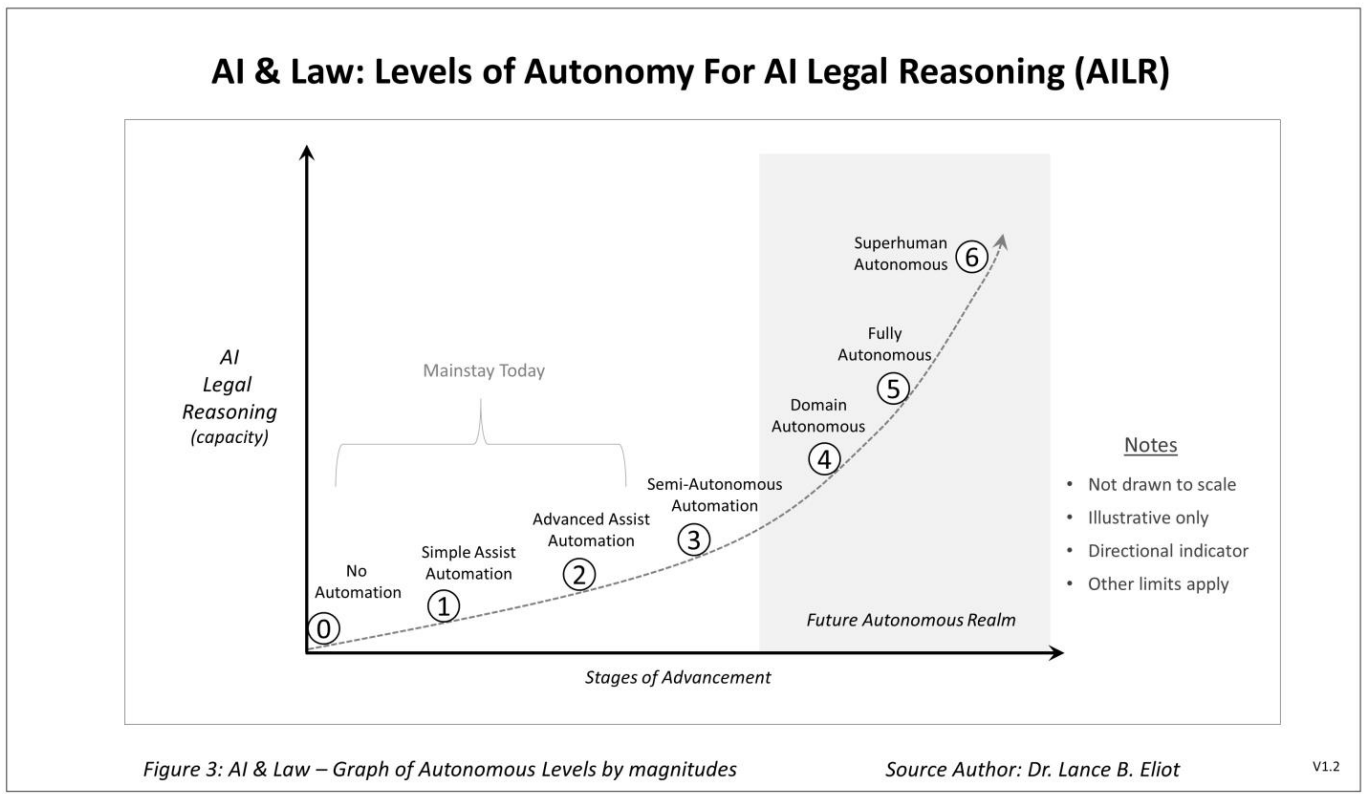

Figure 3: AI & Law – Graph of Autonomous Levels by magnitudes    Source Author: Dr. Lance B. Eliot    V1.2